# Topological Hall effect in single thick SrRuO$_3$ layers induced by defect engineering

Changan Wang,*,1,2,3 Ching-Hao Chang,4 Andreas Herklotz,5 Chao Chen,6 Fabian Ganss,7 Ulrich Kentsch,1 Deyang Chen,6 Xingsen Gao,6 Yu-Jia Zeng,*, 3 Olav Hellwig,1,7 Manfred Helm,1, 3 Sibylle Gemming,1,7 Ying-Hao Chu,8 and Shengqiang Zhou *,1

1) Helmholtz-Zentrum Dresden-Rossendorf, Institute of Ion Beam Physics and Materials Research, Bautzner Landstr. 400, 01328 Dresden, Germany

2) College of Physics and Optoelectronic Engineering, Shenzhen University, 518060 Shenzhen, China

3) Technische Universität Dresden, D-01062 Dresden, Germany

4) Department of Physics, National Cheng Kung University, Tainan 70101, Taiwan

5) Institute for Physics, Martin-Luther-University Halle-Wittenberg, 06120 Halle, Germany

6) Institute for Advanced Materials and Guangdong Provincial Key Laboratory of Optical Information Materials and Technology, South China Academy of Advanced Optoelectronics, South China Normal University，Guangzhou 510006, China

7) Institute of Physics, Chemnitz University of Technology, D-09107 Chemnitz, Germany

8) Department of Materials Science and Engineering, National Chiao Tung University, Hsinchu, Taiwan

---

* Author to whom correspondence should be addressed. e-mails: changan.wang@hzdr.de , yjzeng@szu.edu.cn, and s.zhou@hzdr.de

**ABSTRACT**

The topological Hall effect (THE) has been discovered in ultrathin SrRuO$_3$ (SRO) films, where the interface between the SRO layer and another oxide layer breaks the inversion symmetry resulting in the appearance of THE. Thus, THE only occurs in ultra-thin SRO films of several unit cells. In addition to employing a heterostructure, the inversion symmetry can be broken intrinsically in bulk SRO by introducing defects. In this study THE is observed in 60 nm thick SRO films, in which defects and lattice distortions are introduced by helium ion irradiation. The irradiated SRO films exhibit a pronounced THE in a wide temperature range from 5 K to 80 K. These observations can be attributed to the emergence of Dzyaloshinskii-Moriya interaction as a result of artificial inversion symmetry breaking associated to the lattice defect engineering. The creation and control of the THE in oxide single layers can be realized by *ex situ* film processing. Therefore, this work provides new insights into the THE and illustrates a promising strategy to design novel spintronics devices.

Electron transport coupled with magnetism has attracted a great deal of attention due to its exotic physical properties and potential applications for developing high density memory devices.[1-4] Among different magneto-transport phenomena, the ordinary Hall effect is the result of the Lorentz force and thus is proportional to the magnetic field. The anomalous Hall effect (AHE) is related to magnetization in ferromagnets, which generally stems from a Berry phase in momentum space.[3] A new type of Hall effect, however, is found to depend on neither magnetic field nor magnetization. It originates from the scalar spin chirality $\chi_{ijk} = S_i \cdot (S_j \times S_k)$ generated by non-coplanar or non-collinear spin configurations such as helices, domain walls or skyrmions.[3, 5, 6] When conduction electrons pass through a non-coplanar spin texture, a quantum-mechanical Berry phase is produced in the real space, associated with a fictitious magnetic field. This field is the origin of this particular kind of Hall effect termed as topological Hall effect (THE).[3] The formation of THE are, in most cases, driven by a non-zero Dzyaloshinskii-Moriya interaction (DMI), which requires the presence of strong spin-orbit coupling (SOC) and the breaking of inversion symmetry. Hence, THE induced by skyrmions has been first observed in non-centrosymmetric B20 compounds such as MnSi, MnGe, and FeGe.[7-10] THE due to the existence of topological spin arrangement can also be generated in oxides such as double-exchange ferromagnets and frustrated magnets with pyrochlore structure.[11,12] Recently, an interface-driven THE was discovered in $SrRuO_3$-$SrIrO_3$ bilayers.[3,13, 14] The $SrRuO_3$ (SRO) layer is ferromagnetic (FM), whereas the top "heavy metal" $SrIrO_3$ layer is paramagnetic and provides strong SOC. Because of the broken inversion symmetry, the interfacial DMI appears to drive the formation of magnetic skyrmions and thus the occurrence of the THE in such ultrathin bilayers. Besides, THE was measured in ultrathin (3-6 nm) SRO layers where the sizeable DMI is induced by the ionic displacement in the SRO oxygen-octahedra near the $SrTiO_3$ substrate.[15] Lattice distortions can also be realized through the ferroelectric proximity effect at the $SRO/BaTiO_3$ interface.[4] In the above-mentioned scenarios, however, a heterostructure system is necessary and the SRO layer thickness is limited to only a few nanometres to show a measurable THE. Is it possible to artificially break the inversion-symmetry and therefore induce THE in a not only *single* but also *thick* SRO layer?

In this work, we report that ion irradiation can generate THE in thick SRO layers, hence demonstrating that an interface is not necessary for the appearance of THE. The magnitude of THE closely depends on the irradiation fluence. We attribute the origin of THE to defect engineering by ion irradiation, which breaks the bulk inversion symmetry through altering oxygen octahedral rotation and thus produces DMI in thick SRO films.[16-19]

The crystallographic structures of the SRO films before and after the He ion irradiation were checked by XRD, as shown in **Figure 1**A. The pristine SRO film is epitaxially grown on the STO ($a_{sub}$ = 3.905 Å) substrate. The scan around the (002) reflection for pristine SRO shows a single peak that can be attributed to the orthorhombic (O) bulk-like phase of SRO. The relative intensity of O-like peak decreases after irradiation and its position slightly moves to lower angles with increasing fluence, while a second peak appears and continuously shifts to a lower angle. This second peak can be associated with tetragonal (T) SRO. For both O- and T-peaks, the shift to lower angles can be ascribed to the strain induced by defects that expands the out-of-plane lattice parameter. In the region where the strain is large enough, a phase transition from O- to T-phase occurs.[16] The appearance of XRD fringes at lower angles for the highly irradiated sample is due to the gradient distribution of defects shown in **Figure 1**C, D. This is phenomenon has been observed and explained before, for example, in the study by Boulle and Debelle.[20] The O-T phase transition is also clearly observed in reciprocal space mappings (RSM) around the pseudocubic (103) and (-103) reflections of the pristine and $1\times10^{15}$ film, respectively, as seen in **Figure 1**B. For the pristine sample, single film peaks with different L value are visible. This indicates that the film is single O-phase. The RSMs of the $1\times10^{15}$ film exhibit two peaks associated with the coexistence of O-SRO and T-SRO: the O-SRO peaks have different L values and the T-SRO peaks have the same L value. [16,21] Note that T-SRO and O-SRO all show the same H value as STO, indicating that both phases are still epitaxially locked to the substrate.[22]

We can draw the following picture about the structural evolution of the SRO film with He irradiation. The pristine film is of pure orthorhombic structure, as has been reported for films grown on STO substrates. The orthorhombic phase has an $a^-a^-c^+$ oxygen octahedral rotation pattern.[23] Ion irradiation creates an inhomogeneous defect concentration throughout the film as shown in **Figure 1**C by the stopping and range of ions in matter (SRIM) simulations.[24] The defects induce lattice distortions, most notably a lattice expansion along the perpendicular direction and changes in the oxygen octahedral rotation pattern, as shown in **Figure 1**D, that lead to a transition to a tetragonal structure with $a^0a^0c^+$ octahedral rotations above a certain defect concentration threshold.[16] Consequently, the T-SRO part of the film is getting more intense at the expense of the O-SRO phase as a bigger part of the film is transformed with increasing ion fluence.

**Figure 2** shows the magnetic field dependent Hall resistivity of the SRO films with different He fluences at various temperatures. Red and black curves represent different sweep directions of the magnetic field. For the pristine sample, we can clearly see the conventional behavior of the Hall resistivity, which is dominated by the AHE. For the irradiated samples, however, an additional contribution to the Hall resistivity is observed on top of the AHE loop. For example, for the $1\times10^{15}$ sample the curve for increasing magnetic field at 5 K shows a bump of negative sign between the field strength 0.9 T and 2.4 T. When we sweep along the opposite direction, the dark curve is monotonic in the same field range. Generally, the Hall resistivity is described by

$$\rho_H = R_0 H + R_s M + \rho_H^T$$

where the first, second, and third terms represent the ordinary Hall effect, AHE and THE, respectively. To estimate the contribution of the THE to the Hall resistivity, the first term is already subtracted from the data in **Figure 2**. We can clearly see that the bump behavior is not attributed to the magnetization (M-H) curve (All the films show typical out-of-plane magnetic hysteresis loops, see Figure S1 in Supporting Information). It should be assigned to the third term THE. In order to understand this behavior more clearly, we present a magnified view of the detailed temperature-dependent Hall resistivity of the $1\times10^{15}$ sample in **Figure 3**A. When the temperature is less than 80 K, the Hall resistivity consists of AHE and an additional THE. AHE changes its sign at around 90 K, at which THE also vanishes in the Hall resistivity. Similar phenomenon can also be observed for the $2.5\times10^{15}$ He/cm$^2$ sample (see **Figure 2**). Even for the highest fluence of $5\times10^{15}$ He/cm$^2$, THE is still present at low temperatures, although significantly smaller than for the films with intermediate fluence. At higher temperatures the THE is negligible and we only observe conventional AHE for this sample.

Note that the AHE varies non-monotonically with temperature and the irradiation fluence, and even changes its sign. Similar observation was reported for SrRuO$_3$ and Sr$_{1-x}$Ca$_x$RuO$_3$.[25-27] A scaling behavior between the transverse conductivity σ$_{xy}$ and the magnetization was obtained, which can be well explained by considering the intrinsic origin of the anomalous Hall effect in SrRuO$_3$.[25-27] In the Supplementary Materials, we followed the procedure in Refs. 25-27 to calculate σ$_{xy}$. We obtained a qualitative similar scaling behavior as shown in Figure S2. This also provides strong evidence for the persistence of intrinsic AHE in ion irradiated SRO films.

To precisely investigate the THE, we have extracted it from the Hall resistivity by further subtracting the AHE on basis of field-dependent magnetization data.[28] The procedure is shown in **Figure 3**B using representative data of the $1\times10^{15}$ sample at 60 K. At high fields, the saturation of the Hall resistivity is purely determined by the AHE and the saturation of the magnetization. The residual THE in Figure 3B displays sharp extrema at $\mu_0H = \pm0.16$ T. This indicates that a topological spin structure with scalar spin chirality is introduced in the ferromagnetic SRO matrix near this magnetic field. Similar THE behavior has been reported in bulk and interfacial magnetic skyrmion systems.[3,7-10]

In the following, to see the global feature of $\rho_H^T$, we present a contour map of the derived $\rho_H^T$ as function of both *T* and *H*, as shown in **Figure 3**C for the $1\times10^{15}$ sample. Obviously, the THE is stabilized over a broad region in the *T-H* plane, persisting from 5 K to 80 K. Moreover, the sharp maximum in THE shows a change of sign from negative to positive with increasing temperature, which is in agreement with previous studies.[8,10] The sign of THE is considered to have close relation with the spin polarization of charge carriers that is very sensitive to the position of Fermi level.[10] Therefore, a small change in the band structure due to the strain or temperature variation may influence the spin polarization, and then invert the sign of THE. Besides, the fact that the peak position of $\rho_H^T$ ($H_P$) is close to the coercive field ($H_C$) suggests again that the THE is driven by topologically nontrivial magnetic structures.

We note that previous studies have reported the emergence of the THE in ultra-thin SRO films, originating from the enhanced DMI and the reduced ferromagnetism at the oxide interface to another layer or the substrate.[2-4,13] In our system, however, it is the first time that THE is artificially created in single and thick SRO films without the need of an interface. The sample with a lower fluence ($1\times10^{15}$ He/cm$^2$) shows robust THE in the wide temperature range from 5 K to 80 K, whereas for the SRO film irradiated with the highest fluence ($5\times10^{15}$ He/cm$^2$), the THE becomes very weak. This phenomenon is different with a previous study that the superposition of the Hall signals from two phases can result in similar Hall resistance.[29] In the case of the superposition of the Hall signals from two phases, the abnormal feature in the Hall resistance is more pronounced in a particular intermediate temperature range, indeed not in the low temperature range. In our case, we have the effect more pronounced at lowest measured temperature and decreased with increasing temperature. Therefore, we propose that the observed THE in our system can be attributed to inversion symmetry breaking due to defect engineering by ion irradiation. On the one hand, the created defects expand the out-of-plane lattice constant, which leads to O- to T-SRO phase transition that is accompanied by a change of the oxygen octahedral rotation pattern.[16] This change of oxygen octahedral rotations at the

O-T phase boundary naturally breaks the inversion symmetry, as illustrated in **Figure 1**D. The situation can be compared with earlier studies on ultrathin SRO films where the shift in the octahedral rotation pattern are induced at oxide interfaces.[30] On the other hand, there is a gradient in the density of defects and therefore also in the strain distribution along the depth (see **Figure 1**C).[23,31] Thus, the local structural distortion associated with the gradient strain profile can also lead to the inversion symmetry breaking.[4,19,32] In the study by Wang et al.[4], the authors have calculated the DMI constant by density functional theory. A sizeable DMI emerges as the lattice is distorted due to the ferroelectric polarization from the interface between $BaTiO_3$ and SRO thin films. The DMI constant increases with the ionic displacement between O and Ru along the [001] axis and reaches a maximum when the displacement is 0.15 Å.[4] In our case, ion irradiation induced effects result in an ionic displacement, probably with a broad range, although we cannot quantify it with the current data. One can expect also sizeable DMI in the irradiated SRO films, leading to the formation of nontrivial magnetic structures, which in turn is responsible for the observed THE in the transport. At the highest fluence, the decrease of the THE suggests the decrease in DMI probably due to the further increase of the ionic displacement or the dramatic lattice damage in the film caused by ion irradiation.

In summary, our results demonstrate the presence of THE in a single-thick SRO layer by He ion irradiation. We attribute these observations to the emergence of DMI as a result of artificial inversion symmetry breaking associated to the lattice defect engineering. The possibility to design defects and strain profiles *ex situ* by ion irradiation provides a new path to explore and control the THE in complex transition metal oxides. Moreover, ion irradiation has been well developed for semiconductor-chip technology. It allows for a lateral pattern down to the nanometre scale by lithography and builds a natural link between nontrivial magnetic spintronics and microelectronics.

**Experimental Section**

The SRO thin films (~ 60 nm) were epitaxially grown on SrTiO$_3$ (STO) single crystal substrate by pulsed-laser deposition (PLD) using KrF ($\lambda$ = 248 nm) excimer laser at the repetition rate of 8 Hz, a substrate temperature of 680 °C and in a 75 mTorr oxygen atmosphere. After deposition, these SRO films were irradiated by 6 keV Helium (He) ions with different fluences from 0 to $5\times10^{15}$ He/cm$^2$. The samples were named as pristine, $1\times10^{15}$, $2.5\times10^{15}$ and $5\times10^{15}$, respectively. The phase purity, crystal structure and epitaxial quality were examined by X-ray diffraction (XRD) scan and reciprocal space mapping (RSM) analysis (PANalytical X'Pert PRO diffractometer) using Cu K$\alpha$ radiation. Transport properties were measured using Van der Pauw geometry with a magnetic field applied perpendicular to the film plane in a Lake Shore Hall measurement system. The magnetization data was collected by a superconducting quantum interference device (Quantum Design, SQUID-VSM) magnetometer.

**Acknowledgements**

We would like to thank Prof. R. Ganesh of The Institute of Mathematical Sciences for their enlightening discussions. This study was supported by the Ion Beam Center at Helmholtz-Zentrum Dresden-Rossendorf. C. A. Wang thanks China Scholarship Council (File No. 201606750007) for financial supports. C.-H.C. acknowledges support from the Ministry of Science and Technology, Taiwan (Grant No. 107-2112-M-006-025-MY3), from the Ministry of Education, Taiwan and from Higher Education Sprout Project, Ministry of Education to the Headquarters of University Advancement at National Cheng Kung University. A.H., S. G and S. Z. acknowledge financial support from the German Research Foundation (Grant Nos. HE8737/1-1, GE1202/12-1, ZH 225/10-1). D. Y. C acknowledges financial support from Guangdong Science and Technology Project-International Cooperation project (Grant No. 2019A050510036) and Guangdong Provincial Key Laboratory of Optical Information Materials and Technology (No. 2017B03031007). This work was also supported by the Shenzhen Science and Technology Project under Grant Nos. JCYJ20180507182246321 and JCYJ20170412105400428 and the National Natural Science Foundation of China (Grant Nos. 91963102 and U1832104).

**References**

[1] N. Nagaosa, Y. Tokura, *Nat. Nanotechnol.* **2013,** *8*, 899.

[2] Y. Ohuchi, J. Matsuno, N. Ogawa, Y. Kozuka, M. Uchida, Y. Tokura, M. Kawasaki, *Nat. Commun.* **2018,** *9*, 213.

[3] J. Matsuno, N. Ogawa, K. Yasuda, F. Kagawa, W. Koshibae, N. Nagaosa, Y. Tokura, M. Kawasaki, *Sci. Adv.* **2016,** *2*, e1600304.

[4] L. F. Wang, Q. Y. Feng, Y. Kim, R. Kim, K. H. Lee, S. D. Pollard, Y. J. Shin, H. B. Zhou, W. Peng, D. Lee, W. J. Meng, H. Yang, J. H. Han, M. Kim, Q. Y. Lu, T. W. Noh, *Nat. Mater.* **2018,** *17*, 1087.

[5] Y. Taguchi, Y. Oohara, H. Yoshizawa, N. Nagaosa, Y. Tokura, *Science* **2001,** *291*, 2573.

[6] I. Lindfors-Vrejoiu, M. Ziese, *Phys. Status Solidi B* **2017,** *254*, 1600566.

[7] A. Neubauer, C. Pfleiderer, B. Binz, A. Rosch, R. Ritz, P. G. Niklowitz, P. Böni, *Phys. Rev. Lett.* **2009,** *102*, 186602.

[8] N. Kanazawa, Y. Onose, T. Arima, D. Okuyama, K. Ohoyama, S. Wakimoto, K. Kakurai, S. Ishiwata, Y. Tokura, *Phys. Rev. Lett.* **2011,** *106*, 156603.

[9] X. Z. Yu, N. Kanazawa, Y. Onose, K. Kimoto, W. Z. Zhang, S. Ishiwata, Y. Matsui, Y. Tokura, *Nat. Mater.* **2011,** *10*, 106.

[10] Y. F. Li, N. Kanazawa, X. Z. Yu, A. Tsukazaki, M. Kawasaki, M. Ichikawa, X. F. Jin, F. Kagawa, Y. Tokura, *Phys. Rev. Lett.* **2013,** *110*, 117202.

[11] Y. Machida, S. Nakatsuji, Y. Maeno, T. Tayama, T. Sakakibara, S. Onoda, *Phys. Rev. Lett.* **2007,** *98*, 057203.

[12] H. Yanagihara, M. B. Salamon, *Phys. Rev. Lett.* **2002**, *89*, 187201.

[13] K.-Y. Meng, A. S. Ahmed, M. Baćani, A.-O. Mandru, X. Zhao, N. Bagués, B. D. Esser, J. Flores, D. W. McComb, H. J. Hug, F. Y. Yang, *Nano. Lett.* **2019**, *19*, 3169.

[14] B. Pang, L. Y. Zhang, Y. B. Chen, J. Zhou, S. H. Yao, S. T. Zhang, Y. F. Chen, *ACS Appl. Mater. Interfaces* **2017,** *9*, 3201.

[15] Q. Qin, L. Liu, W. N. Lin, X. Y. Shu, Q. D. Xie, Z. Lim, C. J. Li, S. K. He, G. M. Chow, J. S. Chen, *Adv. Mater.* **2019,** *31*, 1807008.

[16] A. Herklotz, A. T. Wong, T. Meyer, M. D. Biegalski, H. N. Lee, T. Z. Ward, *Sci. Rep.* **2016,** *6*, 26491.

[17] C. A. Wang, C. Chen, C.-H. Chang, H.-S. Tsai, P. Pandey, C. Xu, R. Böttger, D. Chen, Y.-J. Zeng, X. S. Gao, M. Helm, S. Q. Zhou, *ACS Appl. Mater. Interfaces* **2018,** *10*, 27472.

[18] D. Kan, Y. Shimakawa, *Phys. Status Solidi B* **2018,** *255*, 1800175.

[19] Z. L. Li, S. C. Shen, Z. J. Tian, K. Hwangbo, M. Wang, Y. J. Wang, F. M. Bartram, L. Q. He, Y. J. Lyu, Y. Q. Dong, G. Wang, H. B. Li, N. P. Lu, H. Zhou, E. Arenholz, Q. He, L. Y. Yang, W. B. Luo, P. Yu, *arXiv preprint arXiv:1811.10794* **2018**.

[20] A. Boulle, A. Debelle, *J. Appl. Cryst.* **2010,** *43*, 1046.

[21] W. L. Lu, P. Yang, W. D. Song, G. M. Chow, J. S. Chen, *Phys. Rev. B* **2013,** *88*, 214115.

[22] H. W. Guo, S. Dong, P. D. Rack, J. D. Budai, C. Beekman, Z. Gai, W. Siemons, C. M. Gonzalez, R. Timilsina, A. T. Wong, A. Herklotz, P. C. Snijders, E. Dagotto, T. Z. Ward, *Phys. Rev. Lett.* **2015,** *114*, 256801.

[23] W. L. Lu, W. D. Song, P. Yang, J. Ding, G. M. Chow, J. S. Chen, *Sci. Rep.* **2015,** *5*, 10245.

[24] J. F. Ziegler, M. D. Ziegler, J. P. Biersack, *Nucl. Instrum. Methods Phys. Res., Sect. B* **2010,** *268*, 1818.

[25] Z. Fang, N. Nagaosa, K. S. Takahashi, A. Asamitsu, R. Mathieu, T. Ogasawara, H. Yamada, M. Kawasaki, Y. Tokura, K. Terakura, *Science* **2003,** *302*, 92.

[26] R. Mathieu, A. Asamitsu, H. Yamada, K. S. Takahashi, M. Kawasaki, Z. Fang, N. Nagaosa, and Y. Tokura, Phys. Rev. Lett. **2004**, 93, 016602.

[27] R. Mathieu, C. U. Jung, H. Yamada, A. Asamitsu, M. Kawasaki, and Y. Tokura, Phys. Rev. B **2005**, 72, 064436.

[28] Y. Yun, Y. Ma, T. Su, W. Y. Xing, Y. Y. Chen, Y. Y. Yao, R. R. Cai, W. Yuan, W. Han, Phys. Rev. Mater. **2018**, 2, 034201.

[29] D. Kan, T. Moriyama, K. Kobayashi, Y. Shimakawa, *Phys. Rev. B* **2018,** *98*, 180408(R).

[30] Y. D. Gu, Y.-W. Wei, K. Xu, H. R. Zhang, F. Wang, F. Li, M. S. Saleem, C.-Z. Chang, J. R. Sun, C. Song, J. Feng, X. Y. Zhong, W. Liu, Z. D. Zhang, J. Zhu, F. Pan, *J. Phys. D: Appl. Phys*. **2019**, *52*, 404001.

[31] S. Autier-Laurent, O. Plantevin, P. Lecoeur, B. Decamps, A. Gentils, C. Bachelet, O. Kaitasov, G. Baldinozzi, *EPL* **2010,** *92*, 36005.

[32] D-H. Kim, M. Haruta, H-W. Ko, G. Go, H-J. Park, T. Nishimura, D-Y. Kim, T. Okuno, Y. Hirata, Y. Futakawa, H. Yoshikawa, W. Ham, S. Kim, H. Kurata, A. Tsukamoto, Y. Shiota, T. Moriyama, S-B. Choe, K-J. Lee, T. Ono, *Nat. Mater.* **2019,** *18*, 685.

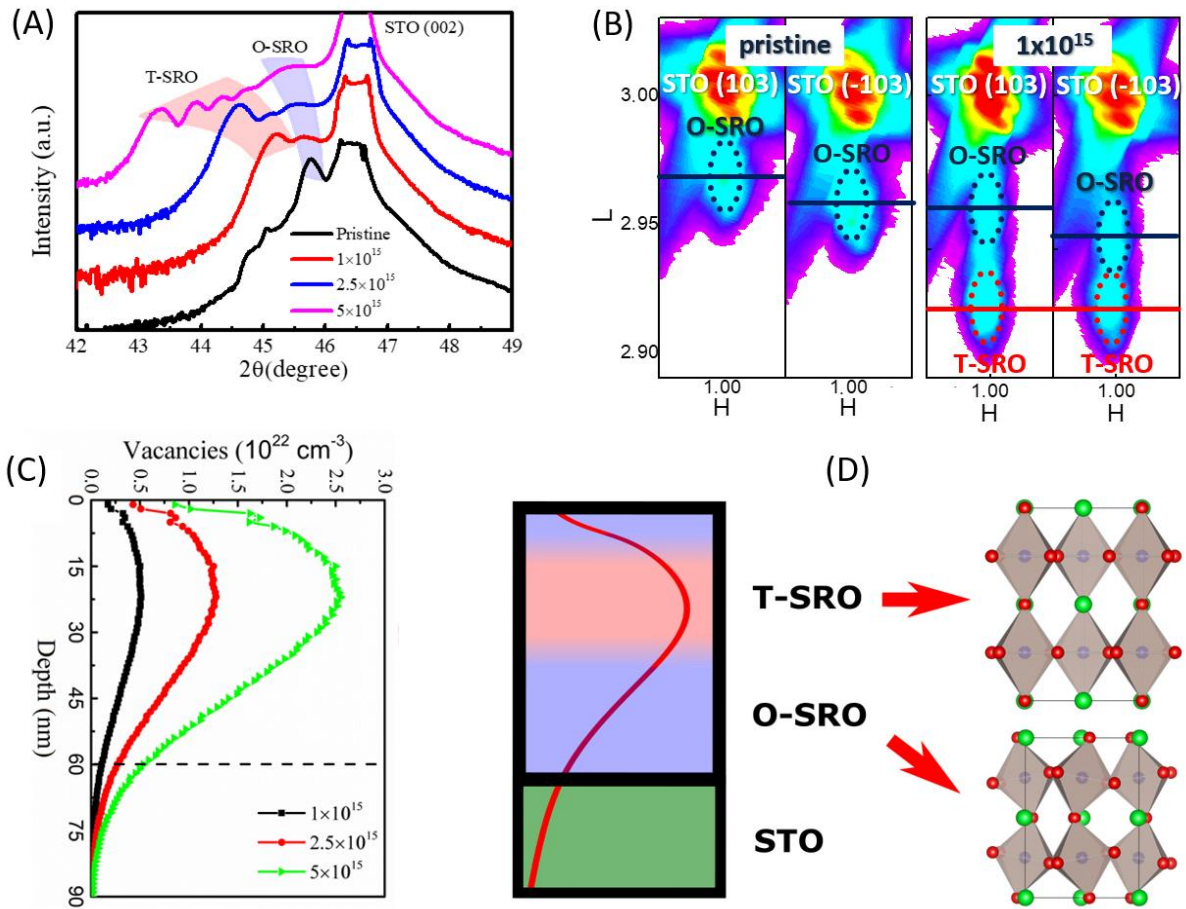

**Figure 1**. Structural characteristics of the SRO films with different helium fluences. (A) 2θ-θ scans around the (002) reflection of SRO films before and after irradiation. The shaded areas highlight the out-of-plane expansion of the O-SRO and T-SRO parts of the film. (B) Reciprocal space mappings (RSM) around the $(103)_{pc}$ and $(-103)_{pc}$ reflections of the pristine and the $1\times10^{15}$ film, respectively. (C) Vacancies vs. depth for different fluences, as obtained from SRIM simulations. The distribution for various fluences is obtained by multiplying the output histogram with the fluence. (D) Corresponding schematic diagram of the structural evolution of the SRO film with He irradiation.

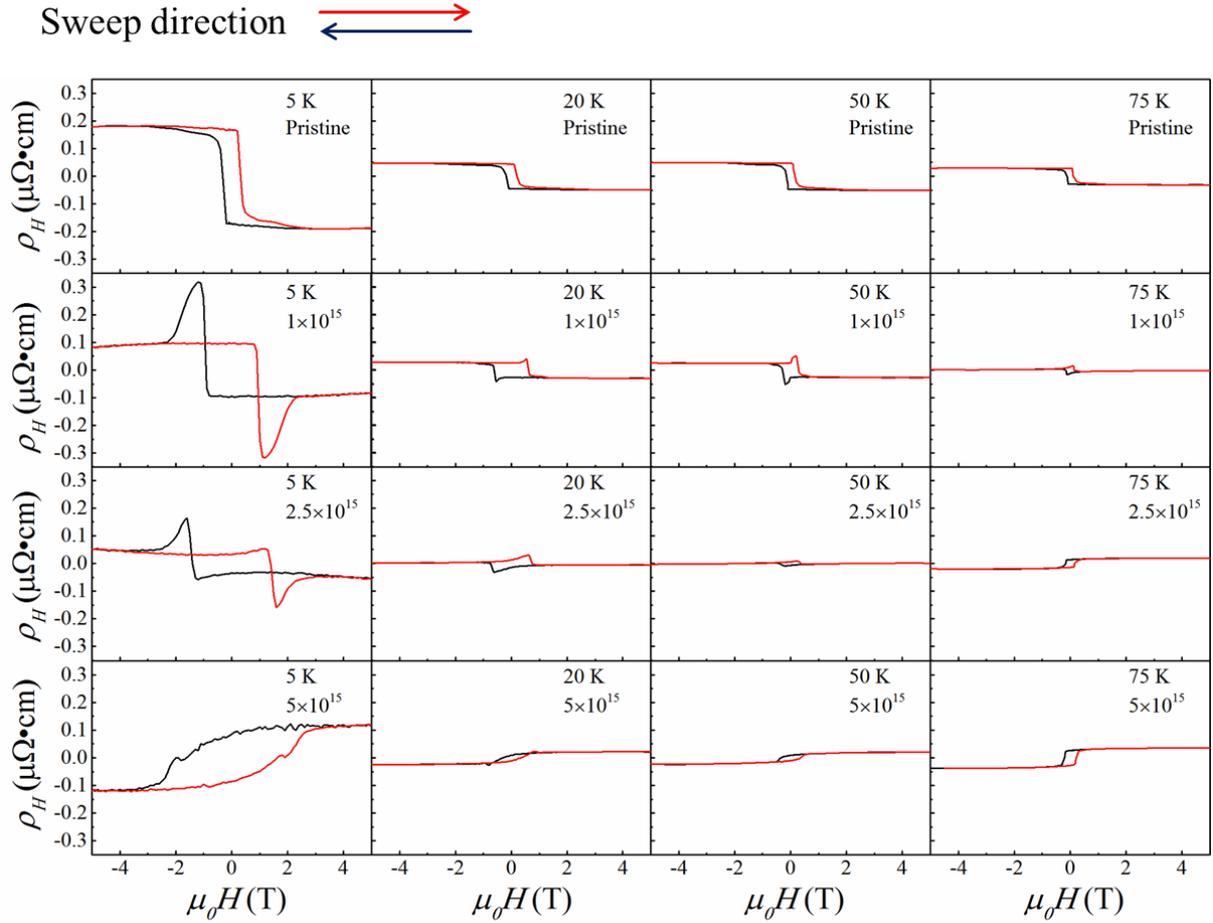

**Figure 2**. Hall resistivity of the samples. Magnetic field dependence of the Hall resistivity ($\rho_H$) of the SRO films with different helium fluences at various temperatures. Red and black represent sweep directions of magnetic field. The ordinary Hall term is subtracted by a linear fit in the high magnetic field region.

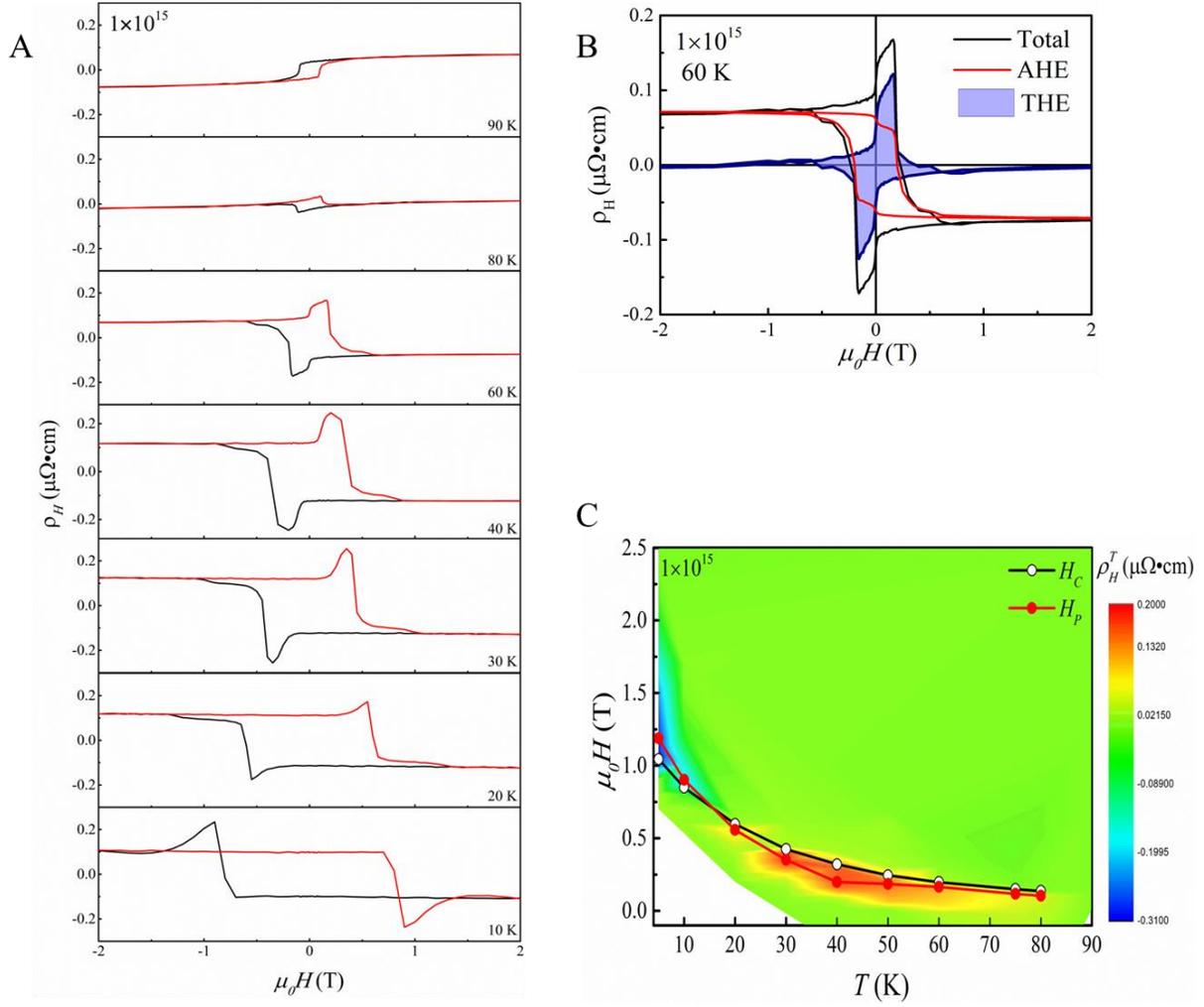

**Figure 3**. (A) Detailed view of the Hall resistivity of the $1\times10^{15}$ sample. (B) Contribution from AHE and THE for the $1\times10^{15}$ sample at 60 K. (C) Color map of topological Hall resistivity in the T-H plane for the $1\times10^{15}$ sample. White open and filled symbols represent coercive field ($H_C$) and the field at which the topological Hall resistivity reaches its maximum ($H_P$), respectively.

Supporting Information

# Topological Hall effect in single thick SrRuO$_3$ layers induced by defect engineering

Changan Wang,*,1,2,3 Ching-Hao Chang,4 Andreas Herklotz,5 Chao Chen,6 Fabian Ganss,7 Ulrich Kentsch,1 Deyang Chen,6 Xingsen Gao,6 Yu-Jia Zeng,*, 3 Olav Hellwig,1,7 Manfred Helm,1, 3 Sibylle Gemming,1,7 Ying-Hao Chu,8 and Shengqiang Zhou *,1

9) Helmholtz-Zentrum Dresden-Rossendorf, Institute of Ion Beam Physics and Materials Research, Bautzner Landstr. 400, 01328 Dresden, Germany

10) College of Physics and Optoelectronic Engineering, Shenzhen University, 518060 Shenzhen, China

11) Technische Universität Dresden, D-01062 Dresden, Germany

12) Department of Physics, National Cheng Kung University, Tainan 70101, Taiwan

13) Institute for Physics, Martin-Luther-University Halle-Wittenberg, 06120 Halle, Germany

14) Institute for Advanced Materials and Guangdong Provincial Key Laboratory of Optical Information Materials and Technology, South China Academy of Advanced Optoelectronics, South China Normal University，Guangzhou 510006, China

15) Institute of Physics, Chemnitz University of Technology, D-09107 Chemnitz, Germany

16) Department of Materials Science and Engineering, National Chiao Tung University, Hsinchu, Taiwan

* Author to whom correspondence should be addressed. e-mails: changan.wang@hzdr.de , yjzeng@szu.edu.cn, and s.zhou@hzdr.de

## Anomalous Hall effect in SrRuO$_3$

Fang *et al.* [1] and Mathieu *et al.* [2, 3] have investigated the anomalous Hall effect in SrRuO$_3$ and Sr$_{1-x}$Ca$_x$RuO$_3$. They observed the Hall resistivity ($\rho_H$) varies non-monotonously with temperature and even changes its sign. There is a scaling behavior between the transverse conductivity σ$_{xy}$ and the magnetization. This can be well explained by their first-principles band calculation by considering the intrinsic origin of the anomalous Hall effect in SrRuO$_3$. By following the calculation procedure in refs. [2, 3], we calculated transverse conductivity σ$_{xy}$. The magnetization was measured for different samples measured at different temperature as shown in Fig. S1. The scaling between σ$_{xy}$ and magnetization is shown in Fig. S2. The scaling behavior is in a qualitative agreement with refs. 1-3. The discrepancy can be due to the appearance of the topological Hall effect in our samples, which adds errors in the extraction of $\rho_{xy}$. This scaling behavior indicates the intrinsic origin of anomalous Hall effect in our SRO films. Ion irradiated SrRuO$_3$ films provide an alternative platform to investigate the intrinsic anomalous Hall effect.

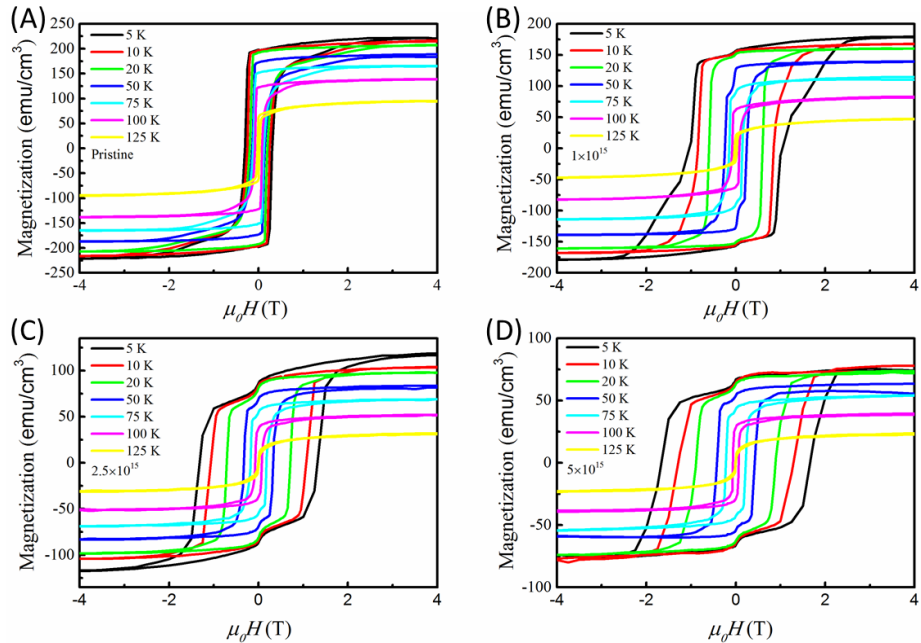

**Figure S1**. Out-of-plane magnetic hysteresis (*M-H*) loops for the SRO films irradiated with different He-ion fluences measured at temperature varying from 5 K to 125 K. All loops display a square-like shape, indicating typical ferromagnetism in SRO films.

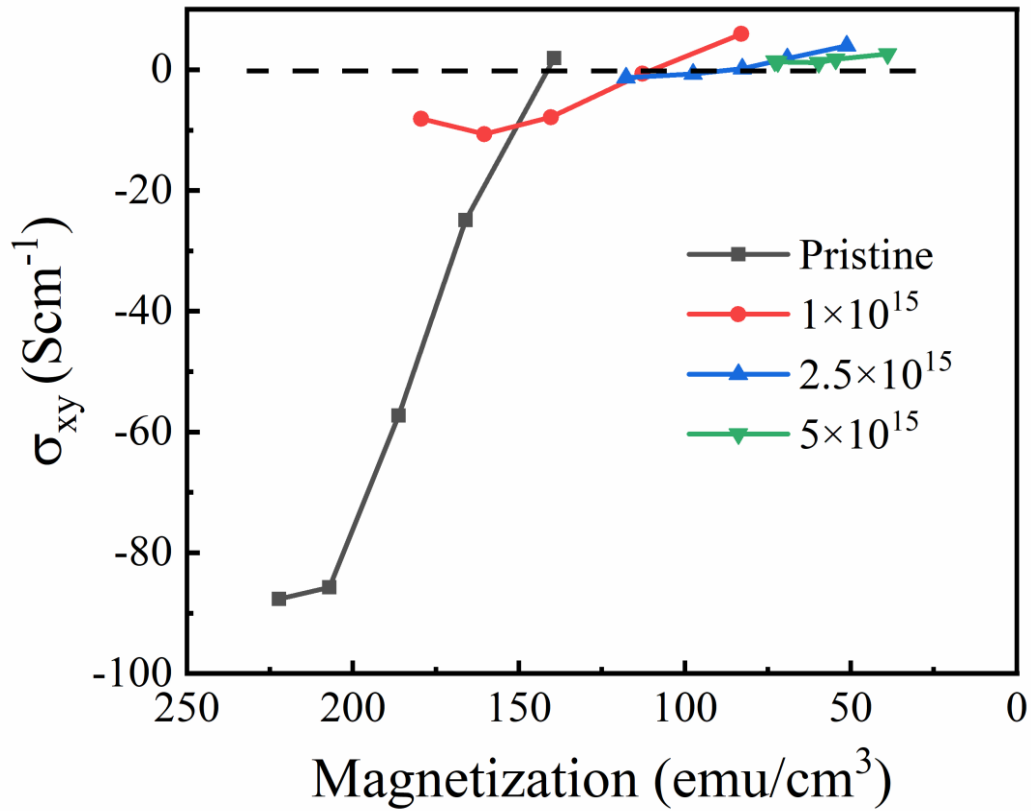

**Figure S2**. The transverse conductivity data obtained for all SRO films vs. the magnetization extracted from Figure S1.

References

[1] Z. Fang, N. Nagaosa, K. S. Takahashi, A. Asamitsu, R. Mathieu, T. Ogasawara, H. Yamada, M. Kawasaki, Y. Tokura, K. Terakura, *Science* **2003,** *302*, 92.
[2] R. Mathieu, A. Asamitsu, H. Yamada, K. S. Takahashi, M. Kawasaki, Z. Fang, N. Nagaosa, and Y. Tokura, Phys. Rev. Lett. **2004**, 93, 016602.
[3] R. Mathieu, C. U. Jung, H. Yamada, A. Asamitsu, M. Kawasaki, and Y. Tokura, Phys. Rev. B **2005**, 72, 064436.